\documentclass[5p,twocolumn,times,number]{elsarticle}

\usepackage{graphicx}
\usepackage{amsmath}   
\usepackage{lineno}

\begin{document}

\begin{frontmatter}

\title{ Cherenkov Detectors Fast Simulation Using Neural Networks}
\author[HSE]{Denis Derkach}
\author[HSE,ROMA]{Nikita Kazeev\corref{cor}}
\ead{Nikita.Kazeev@cern.ch}
\author[HSE,YSDA]{Fedor Ratnikov}
\author[HSE,YSDA]{Andrey Ustyuzhanin}
\author[MIPT,YSDA]{Alexandra Volokhova}

\cortext[cor]{Corresponding author}

\address[HSE]{National Research University Higher School of Economics, Moscow, Russia}
\address[ROMA]{Universit\`{a} di Roma La Sapienza, Roma, Italy}
\address[YSDA]{Yandex School of Data Analysis, Moscow, Russia}
\address[MIPT]{Moscow Institute of Physics and Technology, Moscow, Russia}

\begin{abstract}
 We propose a way to simulate Cherenkov detector response using a generative adversarial neural network to bypass low-level details. This network is trained to reproduce high level features of the simulated detector events based on input observables of incident particles. This allows the dramatic increase of simulation speed. We demonstrate that this approach provides simulation precision which is consistent with the baseline and discuss possible implications of these results.
\end{abstract}



\end{frontmatter}

\section{Introduction}

New runs of the Large Hadron Collider and next generation of colliding experiments with increased luminosity will require an unprecedented amount of simulated events to be produced. This would bring an extreme challenge to the computing resources. Thus new approaches to events generation and simulation of detector responses are needed. Cherenkov detectors, being relatively slow to simulate, are well suited for applying recent approaches to fast simulation. Until recently, the most popular approaches were tabulated response~\cite{Grindhammer:1989zg} and parameterization of detector response~\cite{Rahmat:2011xp}. While both approaches produce valuable results, they require a significant effort at each retuning. That is why a new way to parameterize the detector response needs to be introduced. This way can be paved using a modern day machine learning tools. The most common approach used now is Generative adversarial neural networks (GANs)~\cite{2014arXiv1406.2661G}, while variational autoencoders can also be used~\cite{2013arXiv1312.6114K}. This paper present the first attempt to parameterize a Cherenkov detector response using GANs. 

\section{Fast Simulation Method}
GANs provide a rule to connect input observables with distributions of output ones~\cite{2014arXiv1406.2661G}. A first attempt to apply the GAN to fast simulation in physics analyses was performed recently in~\cite{Paganini:2017hrr}. This attempt used a Geant4 generated calorimeter response as a training sample with the aim to mimic the low-level detector response. 

Our model instead concentrates on the high-level observables reconstruction, thus, effectively bypassing the photon generation stage. This allows us to concentrate on the quality of simulation in terms of the observables used in further analysis. 

Various divergences can be minimised when training GANs. We find that the Wasserstein GAN~\cite{zbMATH03429908} that has already been explored in the context of HEP shows good behaviour. The distance can be written as  a Kantorovich-Rubinstein dual representation~\cite{zbMATH03133048}:
\begin{equation}
    W(p_\theta, p_r) = \sup_{f\in \rm{Lip}_1(X)}\left({\mathbb{E}}_{x\sim p_r}[f(x)] -{\mathbb{E}}_{x\sim\label{eq:dual} p_\theta}[f(x)]\right),\end{equation}
here $p_\theta$ is the optimised parametric distribution, $p_r$ is the empirical distribution. $f$ is an element of 1-Lipshitz function space. This condition allows us to construct a faster converging and more reliable local operator, with $f$ that can be approximated by a sufficiently complex neural network.

However, we also find that the best solution is a Cramer GAN~\cite{2017arXiv170510743B}, which provides unbiased gradients leading to a better fidelity in reproducing distributions. The metrics can also be written out in form of a dual representation~(\ref{eq:dual}), exchanging $f(x)$ to a specific form described in~\cite{2017arXiv170510743B}.

\begin{figure*}[h!]
\centering
\includegraphics[width=0.27\linewidth]{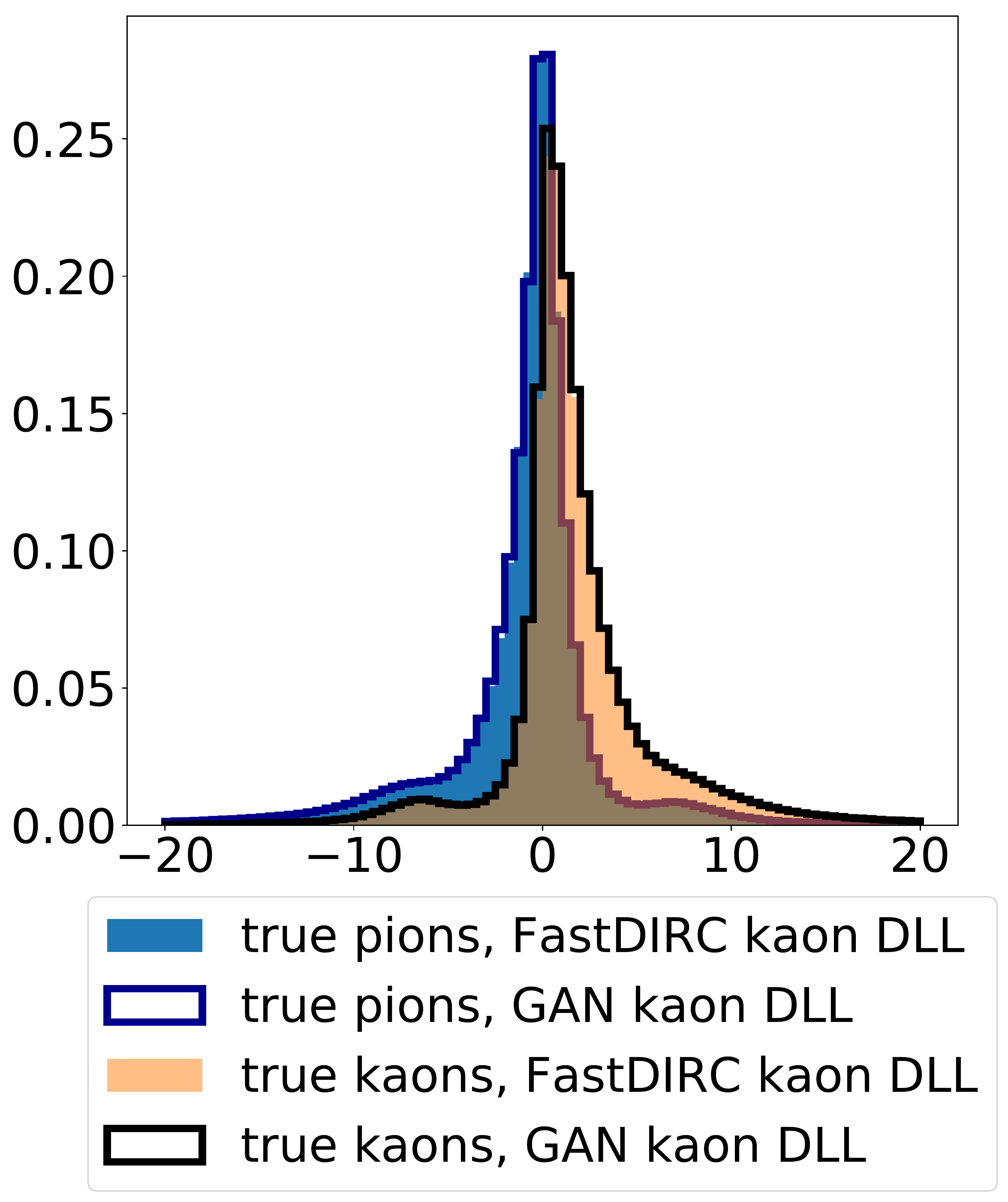}
\includegraphics[width=0.33\linewidth]{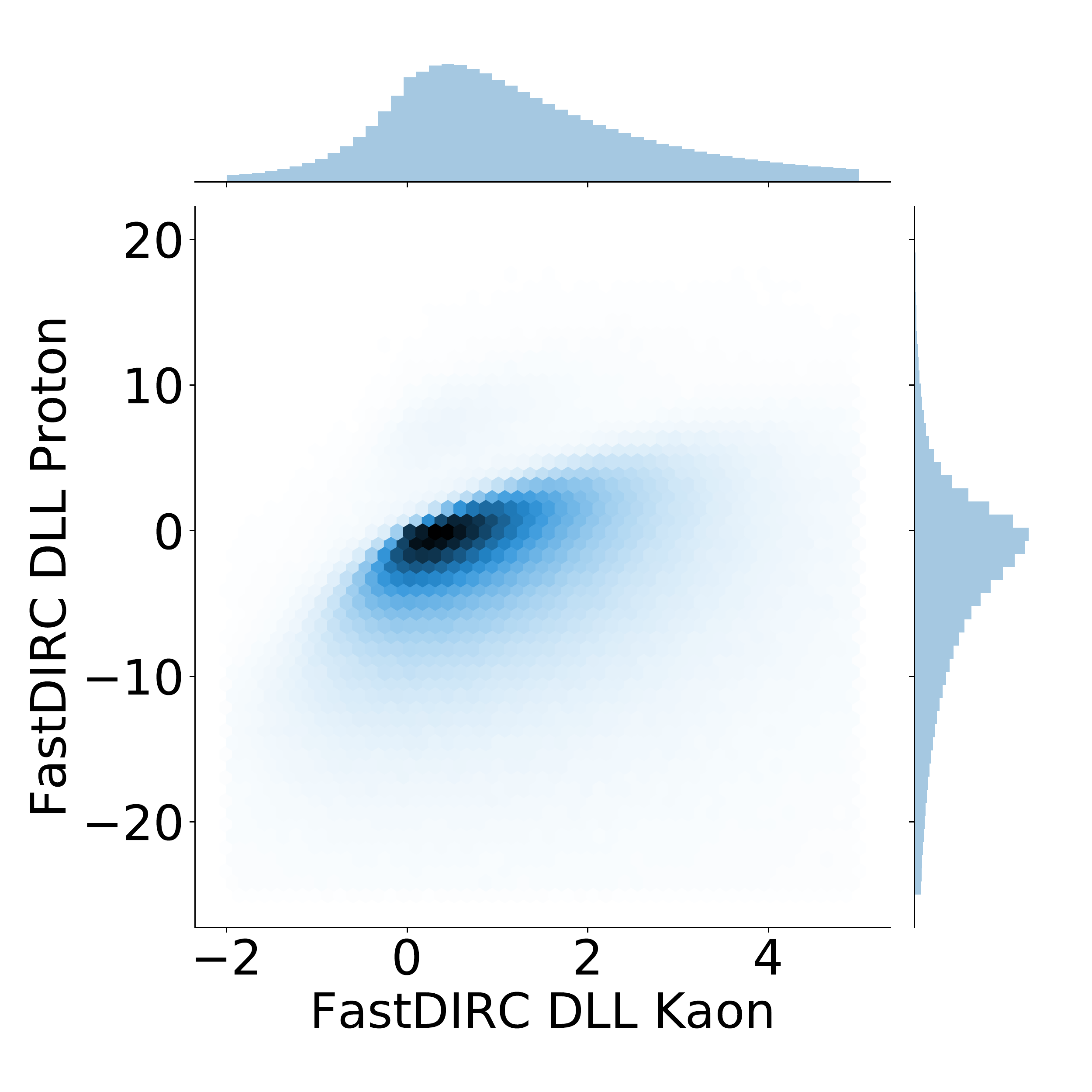}
\includegraphics[width=0.33\linewidth]{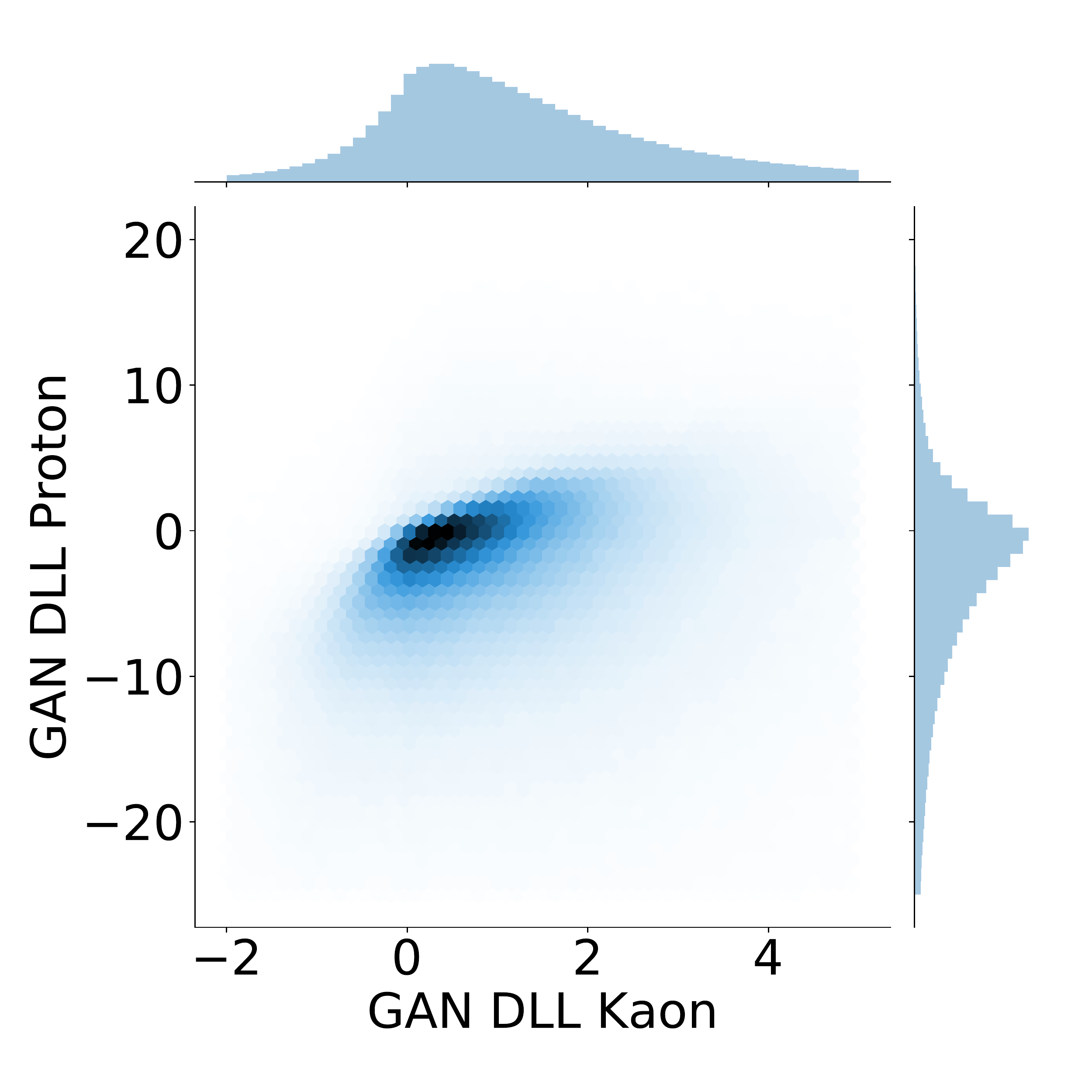}
\caption{Left: An example of 1D projection to kaon delta log-likelihood observables for FastDIRC and GAN simulation for samples consisting true pions (blue) and true kaons (brown). 
Center and Right: An example of 2D projection to kaon and proton delta log-likelihood observables for FastDIRC (left) and GAN (right) simulation. The sample made of true pion.}
\label{fig:LLs}
\end{figure*}
\begin{figure*}[h!]
\centering
\includegraphics[width=0.45\linewidth]{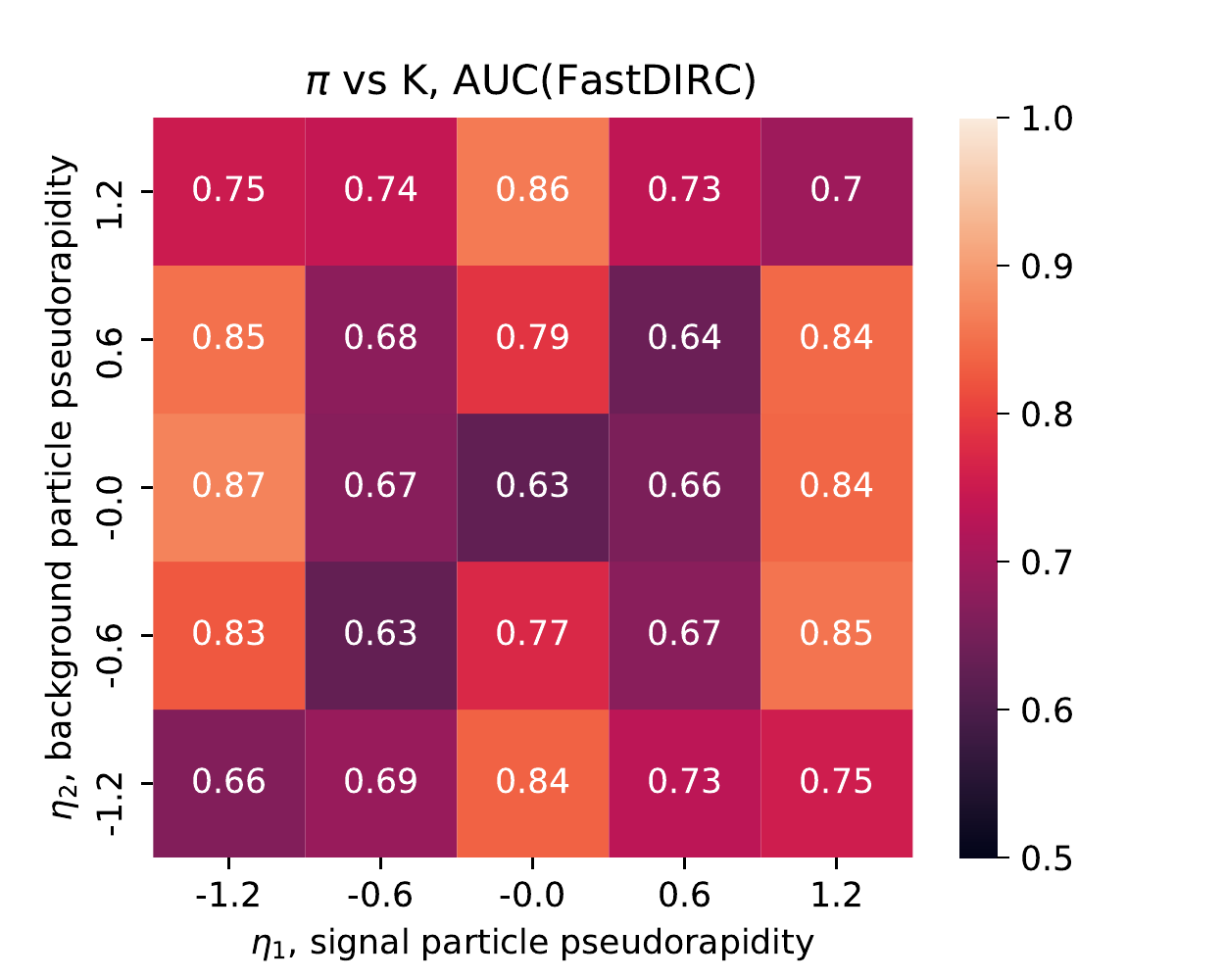}
\includegraphics[width=0.45\linewidth]{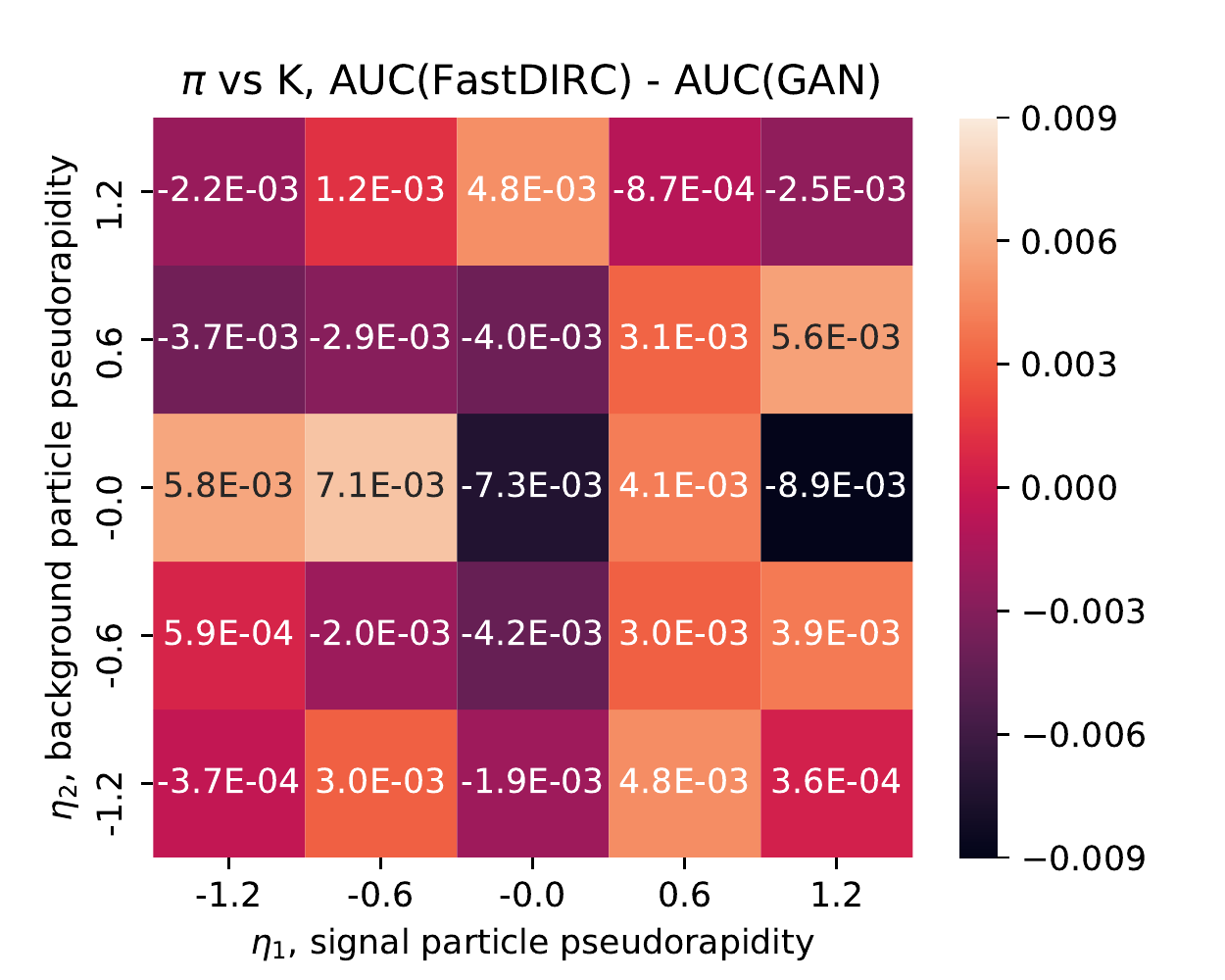}
\caption{Separation power between kaons and pions measured in area under receiver operating characteristic curve score (AUC score). Left is the FastDIRC simulation, right is the difference between GAN and FastDIRC AUC scores. The statistical uncertainty is around 0.005.}
\label{fig:pt-eta}
\end{figure*}
\section{Input data Sample}
As a source of reliable simulated events, we used the FastDIRC~\cite{Hardin:2016cvu} simulation of the Detector of Internally Reflected Cherenkov light. This detector was first used by the BaBar experiment~\cite{Aubert:2001tu} and now is going to be introduced in the GlueX~\cite{Shepherd:2009zz} experiment. The generation is two-fold: in the first stage a sufficient amount of the photons are generated; in the second stage a kernel density estimation is used to produce the likelihood for particle identification. The biggest problem of the fast simulation is expected to come from the interference of two adjacent particles. That is why we modified the code in order to test multiparticle events.  We simulate tracks sampling a flat distribution in pseudorapidity between -1.5 and  1.5, and Gaussian distribution in energy (mean  6~GeV, width 2~GeV) with energy greater than 2.5~GeV. We also update the code adding an additional "background" particle in each event. In the following, the sample generated by the FastDIRC package is referred to as "FastDIRC", our model results is referred to as "GAN".

\section{Model Study}
We construct a neural network for each particle species that takes signal kinematic observables along with a distance to the closest adjacent particle and train it to predict the multidimensional distribution of the likelihoods produced by FastDirc. 

The amount of input observables and the architecture of the neural network was optimised to obtain a subpercent quality of the prediction. The final architecture design is a 10-dense-layers neural network, each containing 128 neurons. The model was trained using 1 million generated events. Here, the input observables used are the full kinematics of event: energy, pseudorapidity and the distance between the particle track and DIRC bar side of signal particle. We trained a separate model for each signal particle type. We transformed each observable distribution into a normal one using quantile transformation before passing them to neural network

We first check that the predictions are consistent with our expectations for one particle tests: we check one- and multidimensional distributions of the output likelihoods in order to understand, whether they are consistent with the output of FastDIRC. We find the histograms to be in a good agreement. An example of the plots is shown in Fig.~\ref{fig:LLs}.

One of the main possible problem of high level observables generation is their interaction with other particles. We take the influence into account by adding an information about the kinematics of the background particle into input observables list. This information is checked to bring the maximum needed quality to the model. In general, this information can be removed if the quality is sufficient. 

We then study the dependence of the generation bias on the closest neighbouring track position. We check this in bins of the kinematic observables for signal and neighbouring particle. In each bin we calculate the separation power between two signal particle species, pions and kaons. We characterise the separation power using area under receiver operating characteristic (ROC AUC). This is done both for likelihoods generated by FastDIRC and our GAN. The test is considered successful if the difference between AUCs generated by different methods is small. The results are shown in Fig.~\ref{fig:pt-eta}. We conclude that the closest neighbour information inserted is sufficient to overcome the problem.

The obtained generation model is lightweight. The speed improvement with respect to full simulation in GEANT~\cite{Agostinelli:2002hh} is $8\cdot10^4$ times on a single CPU core. The speed is also improved with respect to the FastDIRC generation, where a factor up to 80 can be achieved. The batch generation on GPU produces up to 1 million track predictions per second.


\section{Conclusion}
We present a novel approach of the fast simulation of Cherenkov detectors. This approach is based on the generative adversarial neural networks and is gives a good precision, while being very fast.

\section*{Acknowledgments}

The research leading to these results has received funding from the Russian Science Foundation under grant agreement n° 17-72-20127.

\bibliography{mybibfile}

\end{document}